\documentclass[aps,prd,twocolumn,preprintnumbers,superscriptaddress,nofootinbib,floatfix]{revtex4-1}

\usepackage{amsmath,amssymb}
\usepackage{bm}
\usepackage{graphicx}
\usepackage{epstopdf}
\usepackage{hyperref}
\usepackage{array}
\usepackage[utf8]{inputenc}
\usepackage{soul}
\usepackage[usenames, dvipsnames]{color}

\usepackage{subfigure}
\usepackage{slashed}
\usepackage{afterpage}
\usepackage{psfrag}
\usepackage{framed}

\enlargethispage{\baselineskip}

\hypersetup{
     colorlinks   = true,
     citecolor    = blue
}

\newcommand{\MPl}{m_{\rm Pl}}
\newcommand{\MS}{M_{\odot}}
\newcommand{\RS}{R_{\odot}}

\newcommand{\tM}{\tilde{M}}
\newcommand{\tV}{\tilde{V}}
\newcommand{\tR}{\tilde{R}}

\newcommand{\mL}{\mathcal{L}}

\newcommand{\lp}{\lambda_\phi}

\newcommand{\rhot}{{\tilde \rho}}

\newcommand{\be}{\begin{equation}}
\newcommand{\ee}{\end{equation}}
\newcommand{\bea}{\begin{eqnarray}}
\newcommand{\eea}{\end{eqnarray}}
\newcommand{\bal}{\begin{align} }
\newcommand{\eal}{ \end{align} }

\renewcommand\({\left(}
\renewcommand\){\right)}
\renewcommand\[{\left[}
\renewcommand\]{\right]}

\newcommand{\exclude}[1]{}

\begin{document}

\title{Dilute and dense axion stars}

\newcommand{\FIRSTAFF}{\affiliation{The Oskar Klein Centre for Cosmoparticle Physics,
	Department of Physics,\\
	Stockholm University,
	AlbaNova,
	10691 Stockholm,
	Sweden}}
\newcommand{\SECONDAFF}{\affiliation{Nordita,
	KTH Royal Institute of Technology and Stockholm University,\\
	Roslagstullsbacken 23,
	10691 Stockholm,
	Sweden}}
\newcommand{\THIRDAFF}{\affiliation{University of Zaragoza,
	P. Cerbuna 12
	50009 Zaragoza,
	Spain}}
\newcommand{\FOURTHAFF}{\affiliation{Department of Physics,
	University of Michigan,
	Ann Arbor, MI 48109, USA}}
\newcommand{\FIFTHAFF}{\affiliation{Center for Theoretical Physics,
	MIT,
	Cambridge MA 02139 USA}}
\newcommand{\SIXTHAFF}{\affiliation{Department of Physics and Origins Project,
	Arizona State University,
	Tempe AZ 25287 USA}}
\newcommand{\SEVENTHAFF}{\affiliation{T. D. Lee Institute and Wilczek Quantum Center,
	Shanghai Jiao Tong University, 
	Shanghai  200240, China}}

\author{Luca Visinelli}
\email[Electronic address: ]{luca.visinelli@fysik.su.se}
\FIRSTAFF
\SECONDAFF

\author{Sebastian Baum}
\email[Electronic address: ]{sbaum@fysik.su.se}
\FIRSTAFF
\SECONDAFF

\author{Javier Redondo}
\email[Electronic address: ]{jredondo@unizar.es}
\THIRDAFF

\author{Katherine Freese}
\email[Electronic address: ]{ktfreese@umich.edu}
\FIRSTAFF
\SECONDAFF
\FOURTHAFF

\author{Frank Wilczek}
\email[Electronic address: ]{wilczek@mit.edu}
\FIRSTAFF
\FIFTHAFF
\SIXTHAFF
\SEVENTHAFF

\date{\today}

\preprint{MCTP-17-20A}
\preprint{MIT-CTP/4949}
\preprint{NORDITA-2017-112}

\begin{abstract}
Axion stars are hypothetical objects formed of axions, obtained as localized and coherently oscillating solutions to their classical equation of motion. Depending on the value of the field amplitude at the core $|\theta_0| \equiv |\theta(r=0)|$, the equilibrium of the system arises from the balance of the kinetic pressure and either self-gravity or axion self-interactions. Starting from a general relativistic framework, we obtain the set of equations describing the configuration of the axion star, which we solve as a function of $|\theta_0|$. For small $|\theta_0| \lesssim 1$, we reproduce results previously obtained in the literature, and we provide arguments for the stability of such configurations in terms of first principles. We compare qualitative analytical results with a numerical calculation. For large amplitudes $|\theta_0| \gtrsim 1$, the axion field probes the full non-harmonic QCD chiral potential and the axion star enters the {\it dense} branch. Our numerical solutions show that in this latter regime the axions are relativistic, and that one should not use a single frequency approximation, as previously applied in the literature. We employ a multi-harmonic expansion to solve the relativistic equation for the axion field in the star, and demonstrate that higher modes cannot be neglected in the dense regime. We interpret the solutions in the dense regime as pseudo-breathers, and show that the life-time of such configurations is much smaller than any cosmological time scale.
\end{abstract}

\maketitle

\section{Introduction}

The QCD axion~\cite{Weinberg:1977ma, Wilczek:1977pj, Shifman1980493, Kim:1980ax, Zhitnitsky:1980tq, Dine:1981rt, Preskill:1982cy, Abbott:1982af, Dine:1982ah} arising within the Peccei-Quinn solution of the strong CP-problem~\cite{Peccei:1977hh, Peccei:1977ur} is one of the best motivated dark matter candidates. Other bosonic dark matter candidates include axion-like particles~\cite{Arias:2012az} emerging in many extensions of the Standard Model, especially in string theory compactifications~\cite{Svrcek:2006yi, Arvanitaki:2009fg, Ringwald:2012cu, Halverson:2017deq}.

If bosons comprise the dark matter of our Universe, they could form dense (with respect to the average dark matter density) clumps called {\it boson stars}.  ~\cite{Kaup:1968zz, Ruffini:1969qy}, or {\it axion stars} in the specific case of axion dark matter. (Here ``star'' is used to denote an object sustained by hydrostatic equilibrium, whether or not it emits light.) Such objects have been long studied~\cite{Kaup:1968zz, Ruffini:1969qy,Das_1963, Feinblum:1968nwc, Teixeira:1975ad,Colpi:1986ye, Seidel:1991zh, Tkachev:1991ka, Kolb:1993zz, Kolb:1993hw}, and recently there has been revived interest~\cite{Braaten:2015eeu, Eby:2015hyx, Eby:2016cnq, Levkov:2016rkk, Helfer:2016ljl, Braaten:2016dlp, Braaten:2016kzc,Bai:2016wpg, Eby:2017xaw, Desjacques:2017fmf}.

In this article, we study the stability of axion stars as a function of the amplitude of the axion field at the core of the star $|\theta_0| \equiv |\theta(r=0)|$. Our results apply to the full range of axion masses for which QCD axions can comprise all of the dark matter. We identify three distinct branches of axion stars, distinguished by the field amplitude at the core, which in turn determines the density of the star. We should keep in mind, that the axion is a periodic field with amplitude effectively restricted to the domain $0 \leq |\theta_0| \leq \pi$.

For small field values $|\theta_0| \lesssim 10^{-6} \left( 10^{-5}\,{\rm eV} / m \right)$ with $m$ the axion mass, the axion field only probes the harmonic part of the potential, and it can be treated as a free field. In this regime, self-gravity is balanced by the kinetic pressure arising from the uncertainty principle. We call this the {\it dilute} axion star branch. We reproduce the previous findings in the literature for the mass-radius relationship, $R \propto M^{-1}$, where $R$ and $M$ are the radius and mass of the star, respectively. In this regime, the configuration is stable against perturbation: For a given mass $M$, stars are pulled back to the equilibrium radius if they expand because then the (attractive) self-gravity is stronger than the (repulsive) kinetic pressure; conversely, if they are perturbed to smaller radii, they expand because kinetic pressure becomes stronger than self-gravity.

For configurations with $M \sim 10^{-11} M_\odot \left( 10^{-5}\,{\rm eV}/m \right)^2$, self-interactions cannot be neglected anymore, although the amplitude is still comparatively small, $|\theta_0| \sim 10^{-6} \left( 10^{-5}\,{\rm eV} / m \right)$. For QCD axions, the lowest order self-interaction is an attractive quartic term. For amplitudes $|\theta_0| \gtrsim 10^{-6} \left( 10^{-5}\,{\rm eV} / m \right)$, the attractive quartic self-interaction is stronger than gravity, which is negligible in this regime. In this {\it critical\/} branch, we find solutions when the quartic self-interaction balance the kinetic pressure with mass-radius relation $R \propto M$. Note, that this relation implies that axion stars become lighter with growing density, such that they always have masses $M \lesssim 10^{-11} M_\odot \left( 10^{-5}\,{\rm eV}/m \right)^2$ in this branch. However, the solutions are unstable against perturbations: for a given mass $M$, stars expand when perturbed to radii larger than the equilibrium value since the quartic self-interactions are weaker than the repulsive pressure.  Eventually the configuration relaxes to the dilute, interaction-free regime described in the previous paragraph. Conversely, if configurations are perturbed to radii smaller than the equilibrium value, the quartic interaction is too strong to be balanced by the pressure and the star collapses to even higher densities. 

It has recently been pointed out, that new stable configurations, called {\it dense\/} axion stars, are obtained when the amplitude of the axion field in the core reaches $|\theta_0| = \mathcal{O}(1)$~\cite{Braaten:2015eeu}. For such amplitudes, the axion field scans the full non-perturbative axion potential, and self-interactions must be taken into account to all orders. Using the assumption that the axion field in the star is coherently oscillating at a single frequency, as commonly used in the literature, we obtain the mass-radius relation $M \propto R^3$, in agreement with Ref.~\cite{Braaten:2015eeu}. However, we find that the single-harmonic approximation, which holds in the branches described above, is not accurate for the dense branch. Using a multi-harmonic expansion, we find that higher harmonics are generated with amplitudes comparable to the fundamental mode's amplitude. Heuristically,  the presence of higher harmonics corresponds to the generation of (relativistic) axions by coalescence processes $n a \to a$. We find that configurations on the dense branch decay via emission of relativistic axions, with lifetimes of order $\tau_{\rm life} \sim 10^3/m$, which are much shorter than any cosmological timescale. 

When $|\theta_0| \gtrsim \mathcal{O}(1)$, axions stars are short lived solutions of the relativistic equation, elsewhere known as {\it oscillons}~\cite{Dashen:1975, Kudryavtsev:1975, Makhankov:1978, Geicke:1984, Zakharov:1986, Gleiser:1994, Copeland:1995fq, Salmi:2012ta, Mukaida:2016hwd}. In the literature, similar objects have also been called {\it pseudo-breathers}~\cite{Hormuzdiar:1999uz}, {\it axitons}~\cite{Kolb:1993hw}, or {\it oscillatons} when driven by gravity~\cite{Seidel:1991zh, Matos:2000, UrenaLopez:2001tw}. Since gravity is negligible in the dense branch, the axion field is described by the Klein-Gordon equation with the QCD chiral potential (the $\chi$-Gordon equation). There is a large but scattered literature on finding solutions to related equations. For example, in one dimension, assuming a cosine potential leads to the Sine-Gordon equation, which admits localized {\it breather\/} solutions that are not harmonic~\cite{Ablowitz:1973fn}, i.e. which feature an infinite collection of higher harmonics. In three dimensions oscillons closely resemble the breather solutions of the one dimensional Sine-Gordon equation, but they differ in that they radiate energy  and thus decay in a finite lifetime, though slowly relative to the ``natural'' timescale set by the inverse mass of the particles.

Justifying and expanding upon this concise summary,  the remainder of this paper is as follows.  In Sec.~\ref{sec:Basic equations} we set out the basic equations.  In  Sec.~\ref{sec:Numerical results} we find numerically stable solutions and provide quantitative results for the dilute and the critical axion star branches. In Sec.~\ref{sec:oscillons} we discuss the dense branch, and analyze the equilibrium and metastability of dense configurations in a relativistic framework. In Sec.~\ref{sec:discussion} we use the mass-radius diagram to sketch a qualitative storyline for axion stars, and in Sec.~\ref{sec:conclusion} we summarize and conclude.

\section{Axion stars}\label{sec:Basic equations}

\subsection{Axion Lagrangian}

The axion results from promoting the flavor-neutral CP violating {\em angle} of the standard model, $\theta$, to a dynamical field~\cite{Wilczek:1977pj, Weinberg:1977ma} in the Peccei-Quinn mechanism~\cite{Peccei:1977hh, Peccei:1977ur}. The canonical normalization of the dynamical angle $\theta(x)$ requires a new energy scale $f$, the axion decay constant, to define the axion field $a(x) = \theta(x) f$. In the following we will refer to both $\theta$ and $a$ as the axion field. The dynamics of the axion field under the influence of gravity are described by the action
\bea
	S &=& \int d^4x\, \sqrt{-g}\, \mL =\nonumber\\
	 &=& \int d^4x \,\sqrt{-g}\left(\frac{1}{2} \left(\partial^\mu a \right) \left(\partial_\mu a\right) - V(a/f)\right),
	\label{eq:lagrangian}
\eea
where the metric $g^{\mu\nu}$ is determined by the Einstein equation for the energy momentum tensor of the axion field $T^{\mu\nu}(a)$. We adopt the axion potential~\cite{DiVecchia:1980yfw,diCortona:2015ldu}, 
\begin{equation}
	V(\theta) = \frac{\Lambda^4}{c_z} \(1-\sqrt{1-4c_z\sin^2(\theta/2)}\),
	\label{Vqcd}
\end{equation}
where $\Lambda^4 \approx (75.5 \rm \,MeV)^4$ is the topological susceptibility~\cite{diCortona:2015ldu, Petreczky:2016vrs, Borsanyi:2016ksw} and $c_z \approx z/(1+z)^2 \approx 0.22$ with the ratio of the up and down quark masses $z=m_u/m_d \approx 0.48$. Note, that the minimum of the potential is at $V(0) = 0$ and the maximum at $V(\pi) = \Lambda^4 \left( 1-\sqrt{1-4c_z}\right) /c_z$. The axion mass $m$ and the quartic coupling constant $\lambda$ are defined through
\bea
	m^2 &=& \frac{1}{f^2}\frac{d^2 V}{d \theta^2}\bigg|_{\theta=0} = \frac{\Lambda^4}{f^2} \!=\! \(57\mu{\rm eV}\frac{10^{11}\rm GeV}{f}\)^2,\\
	\lambda &=& \frac{1}{f^4}\frac{d^4 V}{d \theta^4}\bigg|_{\theta=0}  = -(1-3c_z)\frac{m^2}{f^2}.
\eea
Assuming spherical symmetry and expanding the metric to linear order about flat space yields the line element
\begin{equation}
	ds^2 = g_{\mu\nu} dx^\mu dx^\nu \!=\! \left(1 \!+\! 2\phi\right)dt^2 \!-\! \left(1 \!-\! 2\phi\right) dr^2 \!-\! r^2\,d\Omega^2,
	\label{eq:lin_metric}
\end{equation}
where $\phi$ is the gravitational potential, which satisfies the Poisson equation with energy density $\rho = T^{00}(a)$, and $d\Omega$ is the differential solid angle. In the following, we rescale time and radius as $t \to m t$ and $r \to m r$, respectively, so that the Lagrangian in Eq.~\eqref{eq:lagrangian} reads
\be
	\mL = \Lambda^4\[\frac{\dot{\theta}^2}{2} - \frac{|\theta'|^2}{2} - \tV(\theta)\],
	\label{eq:lagrangian1}
\ee
where a dot indicates a derivative with respect to the rescaled time, a prime indicates a derivative with respect to the rescaled radius, and $\tV(\theta) \equiv V(\theta)/\Lambda^4$. Coupling the Poisson equation with the equation of motion obtained from the Lagrangian density $\mL$ gives
\begin{eqnarray}
	&& \ddot{\theta} = \left(1 \!+\! 4\phi\right)\left(\frac{2 \theta'}{r} \!+\! \theta''\right) \!+\! 4\dot{\phi}\,\dot{\theta} \!-\! \left(1 \!+\! 2\phi\right)\frac{d \tV(\theta)}{d \theta}, \label{eq:releqmotion} \\
	&& \phi'' + \frac{2\phi'}{r} = 4\pi \beta \rhot, \\
	&& \rhot = \rho_{\rm kin} + \rho_{\rm grad} + \rho_{\rm pot} = \frac{\dot{\theta}^2}{2} +\frac{|\theta'|^2}{2} + \tV(\theta) \label{eq:reldensity},
\end{eqnarray}
where $\beta \equiv Gf^2 = (f/\MPl)^2$ with the Planck mass $\MPl=1.221\times 10^{19}$\,GeV. The energy density $\rhot \equiv \rho/\Lambda^4$ is dimensionless, and reaches $\rhot \sim 1$ when $|\theta|\sim \pi$ and the axion potential saturates. In Eq.~\eqref{eq:reldensity}, we denote the contributions to the energy density from the kinetic, gradient, and potential components separately. Note, that the gradient energy is due to the momentum of the axion arising from the uncertainty principle. So far, the only approximation used is that gravity is weak, $\phi \ll 1$.

We anticipate one of the results of this paper, namely that the system can be studied in two different regimes depending on whether the axion field is $|\theta| \ll 1$ (the ``dilute'' and the ``critical'' axion star regimes) or $|\theta| \gtrsim 1$ (the ``dense'' axion star regime). In the dilute and critical regimes, the axions comprising the star are non-relativistic and the tools described in Sec.~\ref{sec:non-relativistic limit} below apply. When $|\theta| \sim 1$, a full relativistic description is needed, as we sketch in Sec.~\ref{non-perturbative solutions}.

\subsection{Non-relativistic (single harmonic) limit} \label{sec:non-relativistic limit}

When the non-relativistic limit applies, the axion mass is the largest energy scale in the problem, so that axion stars oscillate at a frequency very close to the axion mass $m$. Despite non-linear interactions arising from a cosine or a chiral potential precluding axion stars solutions from having one single frequency, for small field configurations $|\theta| \ll 1$, the one-frequency approximation
\be
	\theta = \Theta(r)\cos\(\omega t\),
	\label{eq:ansatz_wfct}
\ee
suffices. Here, $\omega$ is the total energy of a constituent axion, in units of the axion mass. We write $\omega = 1 + \epsilon$, where $\epsilon$ accounts for the contribution from the binding, kinetic and self-interaction energies, while the one accounts for the rest mass energy. In the non-relativistic approximation, we have $|\epsilon| \ll 1$ and $\omega \approx 1$.

We further assume that gravity is a weak effect, so that we can drop all terms containing $\phi$ in Eq.~\ref{eq:releqmotion}, except for the term $2\phi\theta$ which is of the same order as $\ddot{\theta} + \theta = \(1-\omega^2\)\theta \approx -2\epsilon\theta$. We split the potential into a mass term and the self interaction as
\be
	\tV(\theta) = \frac{1}{\Lambda^4}\frac{m^2}{2}a^2 + \frac{V_{\rm self}(\theta)}{\Lambda^4} = \frac{\theta^2}{2} + \tV_{\rm self}(\theta) ~.
	\label{eq:separate_self_potential}
\ee
%so that, 
Inserting the representation in Eq.~\eqref{eq:ansatz_wfct} into Eqs.~\eqref{eq:releqmotion}-\eqref{eq:reldensity} and averaging over the period $2\pi/\omega$, we obtain
\begin{eqnarray}
	\Theta'' + \frac{2\Theta'}{r} &\simeq& 2\(W_1(\Theta) + \phi + \frac{\omega^2 - 1}{2}\)\Theta, \label{eq:schroedinger_chi}\\
	\phi'' \!+\! \frac{2\phi'}{r} &\simeq& 4\pi \beta \rhot, \label{eq:poisson_chi}\\
	\rhot &\simeq&\rhot_{\rm kin} + \rhot_{\rm grad} + \rhot_{\rm pot}. \label{eq:rho_chi}
\end{eqnarray}
In the last expressions, we have introduced the energy density terms
\be
	\rhot_{\rm kin} = \frac{\omega^2}{4}\Theta^2, \quad \rhot_{\rm grad} = \frac{|\Theta'|^2}{4}, \quad \rhot_{\rm pot} = \frac{\Theta^2}{4} + W(\Theta),
	\label{eq:energydensitycomponents}
\ee
and we have defined the effective self-interaction potential and its first derivative through 
\begin{eqnarray}
	W(\Theta) &=& \frac{1}{2\pi}\int_0^{2\pi}\tV_{\rm self}\(\theta\)\,d(\omega t),\label{eq:WALP}\\
	W_1(\Theta) &=& 2\frac{dW(\Theta)}{d\Theta^2}.\label{eq:W1ALP}
\end{eqnarray}
For $|\epsilon| \ll 1$, Eq.~\eqref{eq:schroedinger_chi} is a Schr{\"o}dinger equation for the radial eigenfunction $\Theta$ with eigen-energy $\epsilon$, while the energy density reduces to $\rhot = \Theta^2/2$ since the contributions from the gradient term and self-interactions are negligible.

We stress that our procedure, which involves the average over $2\pi/\omega$ of the equation of motion leads to the same results as what was obtained in Ref.~\cite{Guth:2014hsa}, where the authors neglect the rapidly oscillating terms proportional to powers of $\exp(i\omega t)$. As long as gravity is negligible and the single-harmonic approximation in Eq.~\eqref{eq:ansatz_wfct} holds, Eqs.~\eqref{eq:schroedinger_chi}--\eqref{eq:rho_chi} are valid even for relativistic axions. We anticipate, that for (most of) the dense branch, gravity is indeed negligible but the single harmonic approximation no longer holds.

\subsection{Axion potential}

We expand the expression in Eq.~\eqref{Vqcd} as
\bea
	\tV(\theta) &=& \sum_{h=0}^\infty v_h \cos(h \theta), \label{taylorseries}\\
	v_0 &=& \frac{1}{2\pi}\int_0^{2\pi}\,\tV(\theta) d\theta,\\
	v_{h>0} &=& \frac{1}{\pi}\int_0^{2\pi}\,\tV(\theta)\cos h\theta d\theta. \label{taylorseries2}
\eea
In our numerical calculation, we truncate the sum in Eq.~\eqref{taylorseries} to the first five terms $h \leq 5$. This attains a precision below 1\% with respect to the chiral potential in Eq.~\eqref{Vqcd}; this precision is better than the accuracy of the chiral perturbation theory itself. We slightly modify the coefficients $v_h$ so that the truncated potential shows: I) the same minimum $\tV(0) = 0$, II) the same mass $\tV_{\theta\theta} = 1$, and III) the same quartic coupling $\tV_{\theta\theta\theta\theta}=\lp$ as the full chiral potential in Eq.~\eqref{Vqcd}, where the (negative) quantity $\lp = -(1-3c_z)$ is related to the axion quartic self-interaction constant as $\lambda = \lp(m/f)^2$. The numerical values of the corresponding corrected coefficients are given in Table~\ref{table_coefficients} for $z = 0.48$.
\begin{table}[h!]
\begin{center}
\begin{tabular}{l}
\hline
	$v_0 = \phantom{-}1.30264$\\
	$v_1 = -1.4403$\\
	$v_2 = \phantom{-}0.1692$\\
	$v_3 = -0.0404$\\
	$v_4 = \phantom{-}0.0105$\\
	$v_5 = \phantom{-}0.001636$\\
\hline
\end{tabular}
\caption{The coefficients in the truncated series expansion of the chiral potential in Eq.~\eqref{taylorseries}, after the corrections described below Eq.~\eqref{taylorseries2} and for $z=0.48$.}
\label{table_coefficients}
\end{center}
\end{table}

The effective non-relativistic potential in Eq.~\eqref{eq:WALP} is
\be
	W(\Theta) = \(\sum_h  v_h J_0\(h\Theta\) \)- \frac{\Theta^2}{4},
	\label{eq:W_cosine2}
\ee
where $J_0(x)$ is the Bessel function of the first kind of order zero for the argument $x$. Notice that the cosine potential is recovered in the limit $c_z \to 0$, equivalent to setting $v_0 = 1$, $v_1 = -1$, and all other $v_h$ equal to zero in Eq.~\eqref{taylorseries}. The set of Eqs.~\eqref{eq:schroedinger_chi}-\eqref{eq:rho_chi} has been extensively applied to self-gravitating systems made of bosons. For the case of axions, the free case $W_1(\Theta) = 0$ has been studied in Refs.~\cite{Seidel:1991zh, Widrow:1993qq, Marsh:2015wka} following the seminal work in Refs.~\cite{Kaup:1968zz, Ruffini:1969qy}. The potential expanded to the quartic interactions has been studied in Refs.~\cite{Chavanis:2011zi, Chavanis:2011zm, Guth:2014hsa, Cotner:2016aaq}. Ref.~\cite{Braaten:2015eeu} considers the set of Eqs.~\eqref{eq:schroedinger_chi}-\eqref{eq:rho_chi} with the cosine potential, using the expression for the energy density (in our notation) $\rhot = \Theta^2/2$, instead of our Eq.~\eqref{eq:energydensitycomponents} obtained from the full energy-momentum tensor.  This implicitly neglects  contributions from self-interaction and kinetic energy to the energy density, which sources the gravitational potential. As we show below, those contributions to the energy density affect the results for the ``dense'' branch.

\section{Numerical results in the single harmonic approximation} \label{sec:Numerical results}

\subsection{Axion star branches} \label{sec:branches}

\begin{figure}[h!]
\begin{center}
\includegraphics[width=\linewidth]{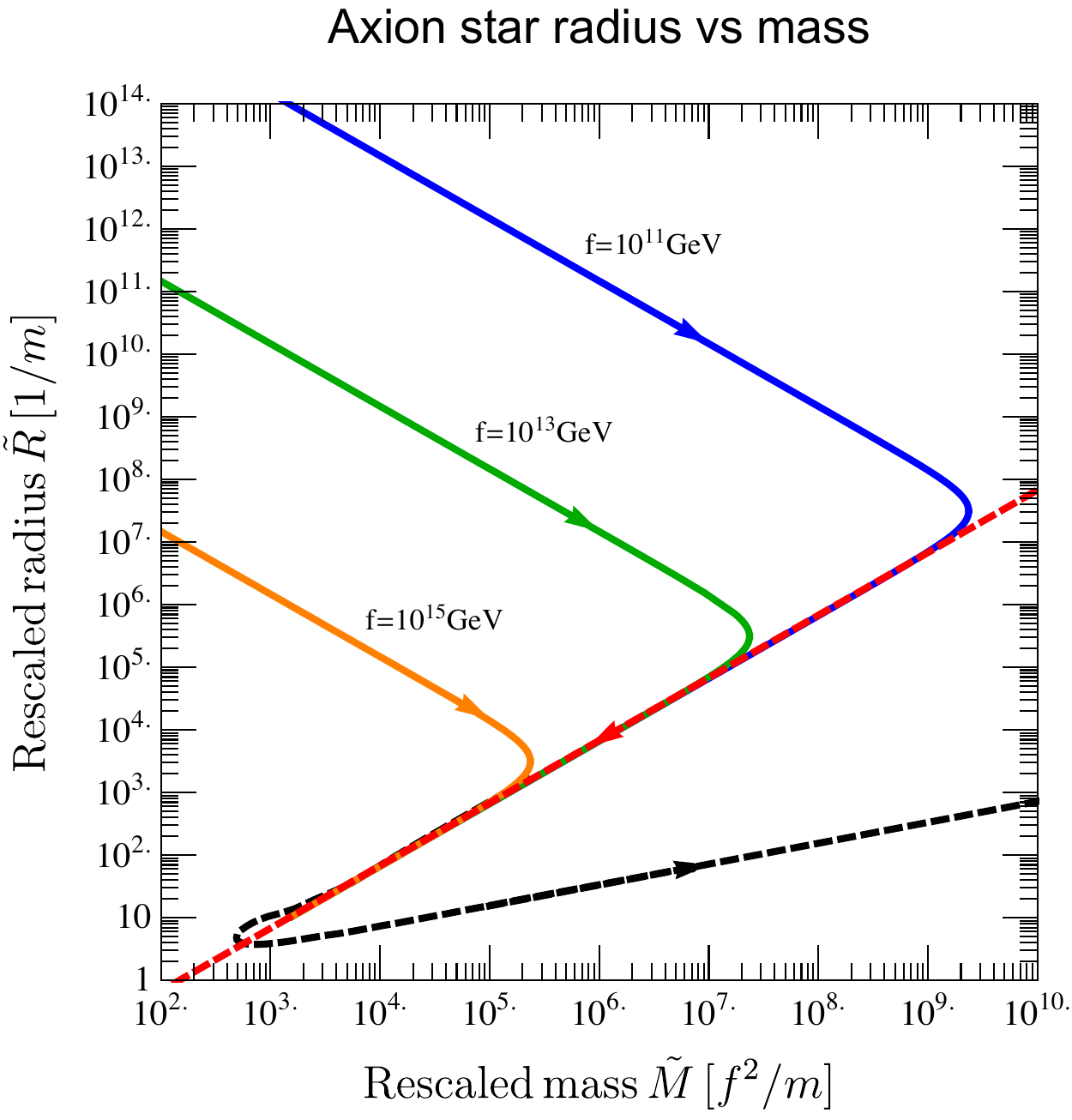}
	\caption{Line of equilibrium solutions of the non-relativistic axion-star equations along the dilute branch for $f=10^{11}\,$GeV (blue), $f=10^{13}\,$GeV (green), $f=10^{15}\,$GeV (orange), connecting to the unstable branch along the critical line (red dashed). Central density increases with the arrows. Also shown is the meta-stable dense solution (dashed black). Note that these results are obtained in the single-harmonic approximation and thus the black dashed curve describing the dense regime should not be trusted.}
	\label{MRdiagram}
\end{center}
\end{figure}

We numerically solve for the radial profile $\Theta(r)$ appearing in the set of Eqs~\eqref{eq:schroedinger_chi}-\eqref{eq:rho_chi}, as a function of the frequency $\omega$. We impose the boundary conditions
\be
	\begin{cases}
	\rhot_0 = (1+\omega^2)|\Theta_0|^2/4+W(\Theta_0), \\
	\Theta(r\to\infty) = 0, \\
	\left.\Theta'\right|_{r = 0} = 0, \\
	\phi(r\to\infty) = 0,
	\end{cases}
\ee
where $\rhot_0$ is the rescaled energy density at $r=0$ and the core amplitude $\Theta_0$ is the amplitude of the axion field at $r = 0$. We obtain a radial profile $\Theta(r)$ via a shooting method, that is by varying the value of the core amplitude $\Theta_0$ until we find a profile that decays as $\exp(-k r)/r$ at a sufficiently large $r$. The solution we seek shows no nodes, and corresponds to the lowest energy state for a given value of $\epsilon$. See Ref.~\cite{Davidson:2016uok} for excited states of an axion star with a quartic potential. We find solutions for all values of $\omega$ within the range (0,1), although the numerics are particularly tricky as we approach $\omega=0$. For each value of $\omega$, we obtain a unique value of the core amplitude and a unique profile. Given the radial profile, we obtain the total mass $M = \int d^3r \rho$ and the radius $R$ of the axion star, the latter defined as the radius containing 90\,\% of the energy~\cite{Ruffini:1969qy}. In Fig.~ \ref{MRdiagram}, we show the mass-radius relation for three values of $ f = \{10^{11}, 10^{13}, 10^{15}\}\,$GeV.\footnote{Note, that for $f = 10^{15}\,$GeV some fine-tuning of the misalignment angle is required to avoid overclosure of the Universe~\cite{Hertzberg:2008wr, Visinelli:2009zm}.} Each point on the line is characterized by a fixed value of $\omega$ and the core amplitude $\Theta_0$. For increasing value of $\Theta_0$, we identify three different regimes: the dilute branch ($|\Theta_0| \lesssim \beta^{1/2}$), the unstable critical configurations ($\beta^{1/2} \lesssim \Theta_0 \lesssim 1$), and the dense branch ($\Theta_0 \gtrsim 1$). For the critical line (red dashed line) and (most of) the dense branch (dashed black line), gravity is negligible.  Then we find universal solutions when expressed in terms of the natural units of star mass, $f^2/m$, and radius, $1/m$. However, gravity is relevant in the dilute branch, where solutions depend on the value of $f$ through $\beta$.

\subsection{Non-relativistic solutions}

In this section, we present heuristic arguments explaining the numerical results obtained in the previous Sec.~\ref{sec:branches} for the dilute and critical branches where $\Theta_0 \lesssim 1$; see also~\cite{Chavanis:2011zi, Chavanis:2011zm} for a similar approach. These branches can be understood in terms of the different contributions to the axion star energy $U$: the gravitational binding energy, the gradient energy, and the (quartic) self-interaction contribution,
\begin{equation} \begin{split}
	U ~&\propto -\frac{GM^2}{R}+\int d^3r\(\frac{f^2}{2}(\theta')^2 + \frac{\lp}{4!}\Lambda^4\,\theta^4\) \\
	&= -\frac{GM^2}{R}+\alpha_k\frac{f^2|\Theta_0|^2}{2R^2 }R^3 + \alpha_4\frac{\lp}{4!}  \Lambda^4 |\Theta_0|^4 R^3 \,.
	\label{eq:potential_energy0}
\end{split} \end{equation}
Here, $\alpha_k$ and $\alpha_4$ are dimensionless parameters which we insert to match the analytical results derived from Eq.~\eqref{eq:potential_energy0} with the numerical solution. Estimating the mass of the axion star as 
\be
	M = \int d^3r \rho \sim  \Lambda^4|\Theta_0|^2 R^3,
\ee
we can express the central amplitude as $|\Theta_0|^2 \sim M/(\Lambda^4 R^3)$,  and the total energy $U$ can be rewritten as
\bea
	U &\propto& -\frac{GM^2}{R} + \alpha_k\frac{f^2M}{2\Lambda^4R^2 } + \frac{\alpha_4\lp}{4!} \frac{M^2}{\Lambda^4 R^3} =\nonumber\\
	&=& \frac{f^2}{m}\[-\frac{\beta \tM^2}{\tR} + \alpha_k\frac{\tM}{2\tR^2 } + \frac{\alpha_4\lp}{4!} \frac{\tM^2}{\tR^3}\].
	\label{eq:potential_energy}
\eea
In the last equality, we have used the scaling property of the Schr{\"o}dinger-Poisson equation, writing the mass and the radius of the star in terms of dimensionless quantities, $\tM = M (m/f^2)$ and $\tR = mR$. The natural scale for the mass and the radius of the axion star are then
\bea
	\frac{f^2}{m} &=& 3\times 10^{-20} \MS \, \(\frac{10^{-5}\,{\rm eV}}{m}\)^3,\\
	\frac{1}{m} &=& 3\times 10^{-11} \RS \, \(\frac{10^{-5}\,{\rm eV}}{m}\),
\eea
where $\MS$ and $\RS$ are respectively the mass and the radius of the Sun. The equilibrium configurations of the axion star can be qualitatively obtained by minimizing the energy density in Eq.~\eqref{eq:potential_energy} with respect to $\tR$, while fixing the axion star mass or, equivalently, the total number of axions $N=M/m$. This gives a quadratic equation whose solutions correspond to the radius of the star for either the dilute branch ($\tR_+$) or the critical branch ($\tR_-$), namely
\be
	\tR_\pm = \frac{\alpha_k}{2\beta \tM}\(1 \pm \sqrt{1 - \frac{\alpha_4|\lp|\beta \tM^2}{2\alpha_k^2}}\).
	\label{eq:mass_radius_relation}
\ee
The stability of the solution is determined by the sign of $\left.\partial^2U/\partial R^2\right|_{R=R_\pm}$. Solutions in the dilute branch ($\rhot_0 \lesssim \beta$) are stable, while those in the critical branch ($\beta \lesssim \rhot_0 \lesssim 1$) are unstable. Matching onto our numerical results from section~\ref{sec:branches}, we obtain
\begin{equation}
	\alpha_k = 9.9, ~~~~~ \alpha_4 = 1.7,
\end{equation}
independent of the value of $\beta$.

The dilute branch of the axion star corresponds to the equilibrium between the gradient energy and gravity. Depending on the value of the decay constant, equilibrium configurations of this type populate the line with negative slope in Fig.~\ref{MRdiagram} with $f = 10^{11}\,$GeV (blue), $f = 10^{13}\,$GeV (green), or $f = 10^{15}\,$GeV (orange), with the mass-radius relation
\be
	\left. \tR_+\right|_{\lp\to 0}  = \frac{\alpha_k}{\beta \tM}.
	\label{eq:dilute_radius}
\ee
For configurations lying above this equilibrium line, the gravitational pull overcomes gradient pressure, so these configurations contract. On the contrary, configurations lying below the mass-radius line in Eq.~\eqref{eq:dilute_radius} are restored to the equilibrium condition by the gradient pressure term. Hence, a restoring force acts to vanish any deviation from the stable equilibrium.

The critical branch, the dashed red line in Fig.~\ref{MRdiagram}, corresponds to the balance of the gradient and the quartic self-interaction energy contributions, with mass-radius relation
\be
	\left. \tR_-\right|_{G\to 0} = \frac{\alpha_4|\lp|\tM}{8\alpha_k}.
	\label{eq:critical_radius}
\ee
Deviations from this configuration are pushed either further towards the dilute branch or to further contraction and are hence unstable. A solution for the radius of the axion star exists as long as the quantity below the square root in Eq.~\eqref{eq:mass_radius_relation} is positive, that is when the mass of the star is smaller than the critical value
\be
	\tM_* = \sqrt{\frac{2\alpha_k^2}{\alpha_4|\lp|\beta}} = \frac{1.3 \times 10^9}{\sqrt{-\lp}} \(\frac{10^{11}\,{\rm GeV}}{f}\),
	\label{eq:mass_critical}
\ee
which corresponds to the radius $\tR_* = $ and to the core amplitude
\bea
	\tR_* &=& \frac{\alpha_k}{2\beta \tM_*} = \sqrt{\frac{\alpha_4|\lp|}{8\beta}}   ,\\
	|\Theta_0^*| &=& \frac{\sqrt{32\beta\alpha_k}}{\alpha_4|\lp|} = \frac{8.8 \times 10^{-8}}{|\lp|}\(\frac{f}{10^{11}\,{\rm GeV}}\).
\eea
The values of $\tM_*$ and $\tR_*$ define the turning point in the top right corner of Fig.~\ref{MRdiagram}, corresponding to the transition from the dilute to the critical branch. In the critical branch, a denser solution corresponds to moving along the red dashed line in Fig.~\ref{MRdiagram} towards the bottom left of the figure, with the star contracting and becoming lighter. Since in this branch the core amplitude increases as $\Theta_0 = \Theta^*_0 M_*/M$, non-perturbative dynamics becomes relevant when $\Theta_0 \approx 1$, or at a typical mass
\bea
	\tM\(\Theta_0 = 1\) &\approx& \Theta^*_0 \tM_* = \(\frac{4\alpha_k}{\alpha_4|\lp|}\)^{3/2} = \frac{110}{|\lp|^{3/2}},
	\label{eq:nonpert_mass}\\
	\tR\(\Theta_0 = 1\) &\approx& \sqrt{\frac{\alpha_k}{\alpha_4|\lp|}} = \frac{2.4}{\sqrt{-\lp}}.
	\label{eq:nonpert_radius}
\eea
These values of $\tM\(\Theta_0 = 1\)$ and $\tR\(\Theta_0 = 1\)$ mark the second turning point in the bottom-left region of Fig.~\ref{MRdiagram}. For larger values of the core amplitude, the axion field explores the whole chiral potential and a different treatment is needed. 

\subsection{Non-perturbative solution} \label{non-perturbative solutions}

The axion star solutions found for $\Theta_0\gtrsim 1$ correspond to a clump of axions whose total mass and radius are larger than the critical values in Eqs.~\eqref{eq:nonpert_mass} and~\eqref{eq:nonpert_radius}. For such configurations, higher order terms in the attractive self-interacting potential cannot be neglected and a new regime is obtained, often referred to as the ``dense'' axion star regime in the recent literature~\cite{Braaten:2015eeu, Braaten:2016dlp}. We show the numerical results for the mass-radius relation obtained in the dense branch configuration with the solid black line in Fig.~\ref{MRdiagram}. Fitting the curve far from the turning point leads to the relation $\tR = 0.6 \tM^{1/3}$. This regime corresponds to classically stable configuration with an almost constant density $\rho \sim \Lambda^4$ in the inner core. For the mass-radius relation, we have obtained the same power-law exponent (1/3) as in Ref.~\cite{Braaten:2015eeu}, because such dependence follows from the fact that the solution in the dense branch saturates the QCD potential and leads to a constant density of the star. 

\begin{figure}[h!]
	\begin{center}
	\includegraphics[width=\linewidth]{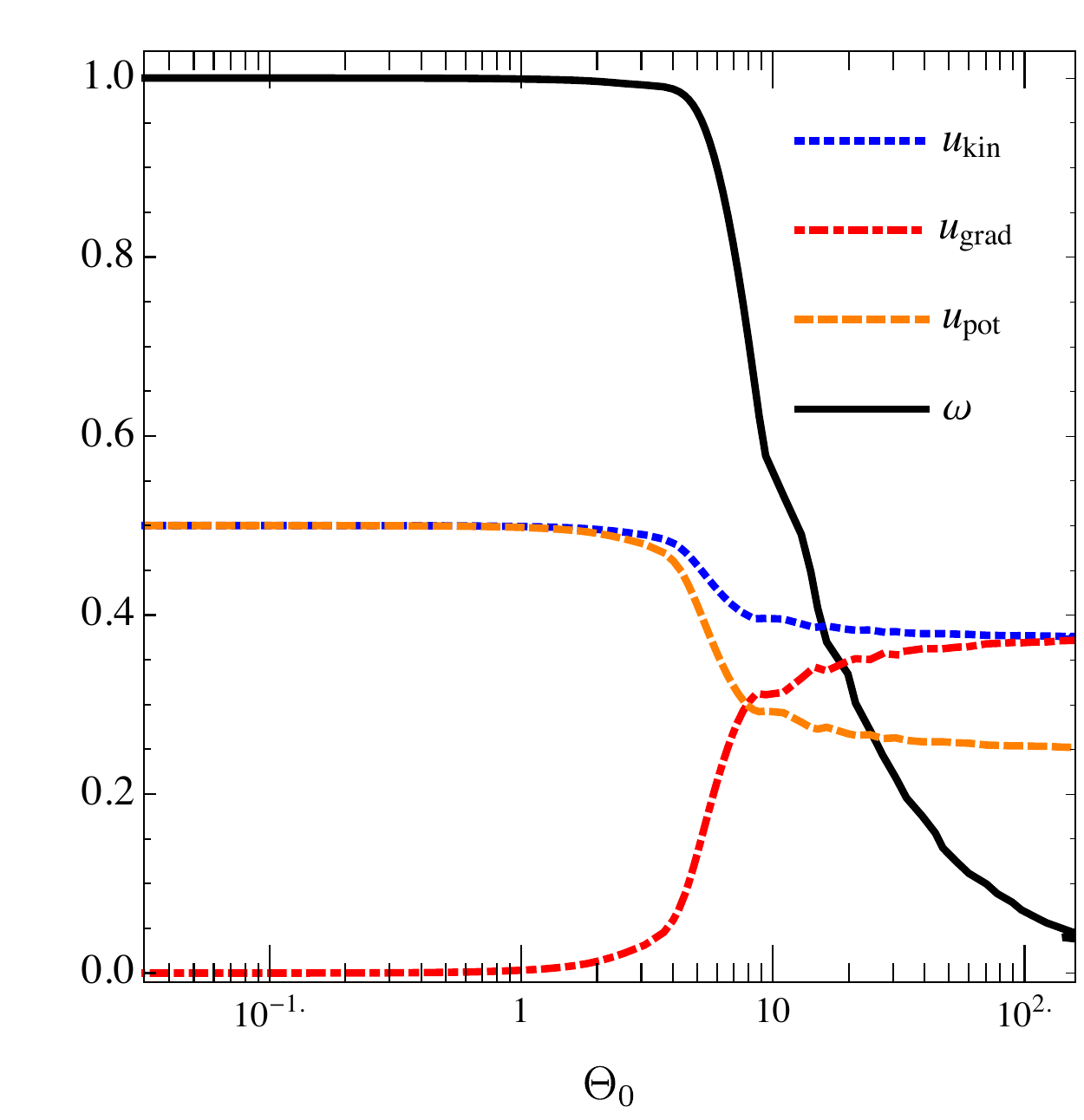}
	\caption{The frequency of the axion star $\omega$ (black solid line) as a function of the core amplitude $\Theta_0$ for our numerical solutions of the non-relativistic stability equations, \eqref{eq:schroedinger_chi}-\eqref{eq:rho_chi}. 
We also show the contributions to the total energy from the kinetic (blue dotted line), gradient (orange dashed line), and potential energy (red dot-dashed line). In the dense branch, i.e. $\Theta_0\gg 1$, the solution is not consistent with the non-relativistic approximation. }
	\label{fig:equipartition}
	\end{center}
\end{figure}

However, the structure of our solution differs greatly from what was obtained in Ref.~\cite{Braaten:2015eeu}.  We disagree on their interpretation of the equilibrium of the axion star in the dense branch for three main reasons. I) We have included the self-interactions and the gradient energy terms through Eq.~\eqref{eq:energydensitycomponents}. These terms cannot be neglected, as we show in Fig.~\ref{fig:equipartition}. II) In Ref.~\cite{Braaten:2015eeu} the set of equations is solved in the Thomas-Fermi approximation, that is neglecting the Laplacian of $\Theta$ appearing on the left-hand side of Eq.~\eqref{eq:schroedinger_chi}. III) Most importantly, the single-harmonic approximation in Eq.~\ref{eq:ansatz_wfct} does not hold in the non-perturbative regime.

\begin{figure}[h!]
	\begin{center}
	\includegraphics[width=\linewidth]{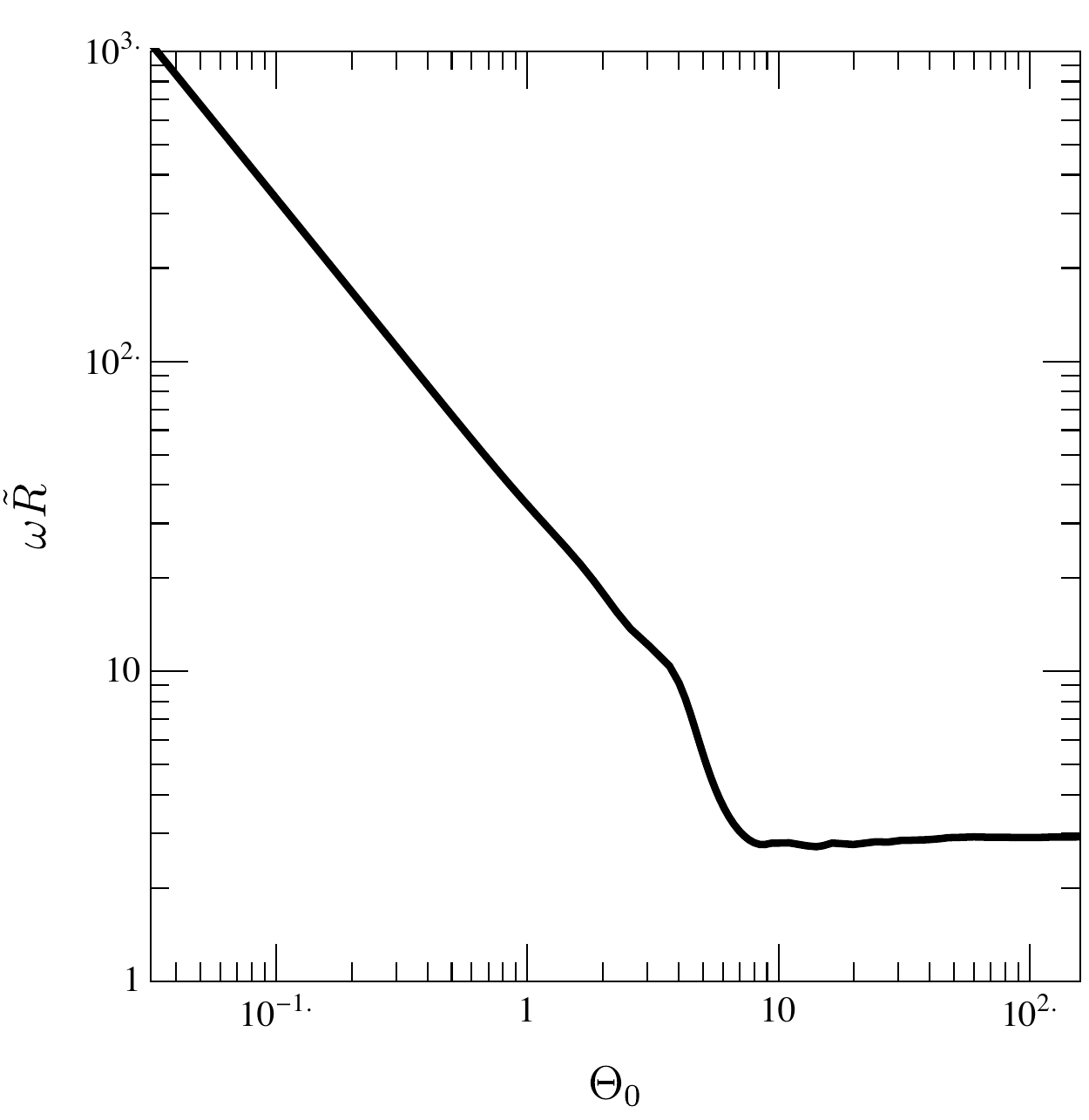}
	\caption{The rescaled axion star radius $\tR$ times the axion frequency $\omega$, as a function of the core amplitude $\Theta_0$.}
	\label{fig:R_omega}
	\end{center}
\end{figure}

In Fig.~\ref{fig:equipartition}, we show the different contributions to the mass of the axion star, $M=\int d^3r \rho$, from the various components in Eq.~\eqref{eq:reldensity}, namely $u_\alpha = \int d^3r \rho_\alpha/M$, where $\alpha \in \[{\rm kin, grad, pot}\]$, as a function of the core amplitude.
In the $\Theta_0 \lesssim 1$ ($\Theta_0 \gtrsim 1$) regime shown, the star is in the critical (dense) branch. In the critical branch, the kinetic and potential energies both contribute a factor equal to $1/2$. This result can be interpreted by the fact that the wave function of the coherent axion field undergoes harmonic oscillations, with the energy density equipartitioned between the kinetic and potential terms. However, as we approach the dense regime, the contribution from the gradient term increases, to the extent that for $\Theta \gtrsim 1$ all three components contribute with a similar magnitude. Thus for dense axion stars the energy density must include all energy contributions.  Also, the Thomas-Fermi approximation is not justified since the Laplacian term is crucial for solving Eq.~\eqref{eq:schroedinger_chi} in the whole domain shown in Fig.~\ref{fig:equipartition} and~\ref{fig:R_omega}.

In more detail, the structure of a dense axion star looks as follows.  The stellar core is composed of relativistic axions since in that region $\omega^2\Theta \sim \nabla^2\Theta$, although self-interactions are not entirely negligible. As we move out of the core, there is an intermediate region where the self-interactions balance the gradient term.  Finally,  in the outmost part self-interactions are again negligible	.

To further illustrate that the axion field is relativistic in the dense regime, in Fig.~\ref{fig:equipartition} we show the axion energy per particle $\omega$ (black solid line), which drops to zero for $\Theta \gtrsim 1$, due to the fact that self-interactions increase with $\Theta_0$. Then, the non-relativistic condition $\omega \gg \pi /\tR$, which expresses that the typical momentum of the axion is much smaller than its energy, no longer holds. Fig.~\ref{fig:R_omega} also shows this conclusion, since the quantity $\omega \tR$ decreases from being much larger than one to a constant value $\sim 3$ for which the non-relativistic interpretation does no longer hold. The inequality $m R \gg 1$, or $\tR \gg 1$, which holds even in the dense branch, is not sufficient to justify a non-relativistic approach.

In addition, our solution shows that gravity is negligible everywhere inside the star. The gravitational energy density at a distance $r$ from the center of the star is $\rho_G = G \rho M(r)/r$, where $M(r)$ is the mass enclosed within the radius $r$, so we can write
\be
	\frac{\rho_G}{\rho} =  \frac{\beta \tM}{\tR} = 4.6\beta \tR^2,
	\label{eq:neglectgravity}
\ee
where in the last step we have used the parametrization $\tR = 0.6 \tM^{1/3}$. Hence, gravity can be neglected for $\tilde R \lesssim \sqrt{1/\beta}$. For $\tR = \mathcal{O}(1)$, gravity can be safely neglected as long as $f \ll \MPl$ or $\beta \ll 1$, which is the range of parameters considered in this work. However, for dense axion star solutions of larger mass, gravity could eventually become important again for $\tR \approx (4.6 \beta)^{-1/2}$. We do not consider this latter possibility here.

As we have previously discussed, the solutions obtained in the dense branch are not self-consistent because the single frequency approximation in Eq.~\eqref{eq:ansatz_wfct} is not justified on the basis of the findings in Fig.~\ref{fig:R_omega}. When the amplitude of the axion field becomes $\Theta = \mathcal{O}(1)$, the axion fields probes the full chiral potential and all orders of self-interaction become relevant. Then, higher harmonic modes of the axion field whose frequency is a multiple of the fundamental mode $\omega = m$ are generated with amplitude comparable to that of the fundamental mode.  In the next section we therefore start over from Eq.~\eqref{eq:releqmotion} and perform a multi-harmonic expansion.

\section{Oscillons} \label{sec:oscillons}

\subsection{Generalities on the relativistic equation}

Based on the findings of the previous Section, axions in the dense regime $\Theta_0 \gtrsim \mathcal{O}(1)$ can be studied using a relativistic approach and ignoring gravity. For simplicity, we derive results for the illustrative case of a cosine potential
\begin{equation}
	V(\theta) = \Lambda^4\left(1-\cos\theta\right),
	\label{eq:Vcos}
\end{equation}
obtained from the chiral potential Eq.~\eqref{Vqcd} for $c_z \to 0$. In that case, the relativistic equation of motion is the Sine-Gordon equation
\be
	\ddot \theta - \theta''-\frac{2}{r}\theta' +\sin\theta = 0.
	\label{eq:sinegordon}
\ee
We wish to identify the oscillon solutions of Eq.~\eqref{eq:sinegordon}, namely the solutions that are spatially-localized and time-periodic. Such solutions circumvent Derrick's theorem~\cite{Derrick:1964}, which states that the scalar field Lagrangian in Eq.~\eqref{eq:lagrangian} expressed in flat space-time does not admit time-independent, finite energy solutions because shrinking a non-zero field configuration effectively reduces the total energy of the system~\cite{Rosen:1968, Lee:1975iw, Friedberg:1976az, Friedberg:1976ay, Coleman:1985}.   Although the {\it ansatz\/} we used previously, Eq.~\eqref{eq:ansatz_wfct}, is not a proper solution for the non-time averaged potential, we expect it to be a reasonable approximation at the transition from the non-relativistic to the relativistic domain when $\Theta_0 \sim 1$. 

There is a long history of searching for oscillons of the Sine-Gordon equation, with the most positive outcome being solutions that last $\mathcal{O}(100-1000)$ oscillations in units of $1/m$~\cite{1976ZhPmR..24...15B, Gleiser:1994, Gleiser:1999tj, Fodor:2006zs, Salmi:2012ta}. The general consensus is that absolutely stable solutions do not exist, although we know of no definite proof.  In any case, is much that we can learn about unstable oscillons from the literature. 

For axions in particular, Kolb and Tkachev~\cite{Kolb:1993hw} discovered the so called ``axitons'' when studying the cosmological evolution of the axion field in the dark matter context. They followed the evolution of the Sine-Gordon equation in an expanding Universe in which the axion mass strongly depends on the cosmic time, and identified an instability condition that leads to small clumps of the axion field with large values $\theta \sim \pi$ to disappear in bursts of relativistic axions. This instability, which originates from the attractive quartic self-interaction term, is well known in the condensed matter community and has been recently revisited in Ref.~\cite{Levkov:2016rkk}. In that paper, the authors follow the collapse of a dilute axion star with a mass slightly above the critical value $M_*$. The axion star solution shows a self-similar collapse that ends when the central amplitude saturates the axion potential. Then, the axion field oscillates for a few times, radiating relativistic axions and relaxing to a small amplitude which is nevertheless larger than the starting value. Such instabilities are triggered for a few times until the central amplitude relaxes to the stability region described above. The simulations include gravity, so that the final state can still be a dilute axion star, but the dynamics of the collapse and the radiation of relativistic axions happens at very small radii where gravity is negligible compared with the self-interactions and gradients. 

The simulations in Ref.~\cite{Levkov:2016rkk} are of considerable phenomenological interest, since in principle the collapse of dilute stars is the most natural mechanism to produce dense axion stars. However, one can address the question of dense axion star stability separately from their possible cosmological origin.   For such a task we need other means.  A promising approach emerged in Ref.~\cite{Piette:1997}, where the authors convert the Sine-Gordon equation into a series of equations with different harmonics.     

\subsection{Beyond the 1st harmonic approximation}

A general time-periodic solution can be written in terms of an infinite numerable set of harmonics. 
Thus we can write our oscillon ansatz as $\theta$ as~\cite{Alfimov:2000}
\be
	\theta = \sum_n \Theta_{2n+1}(r) \cos\[ (2n+1)\omega t\],
	\label{eq:ansatz_fourier}
\ee
which, once plugged into the Sine-Gordon Eq.~\eqref{eq:sinegordon}, yields a set of coupled equations 
for the different harmonics, 
\be
	\begin{cases}
	\Theta''_1 + \frac{2}{r}\Theta'_1 + \omega^2\Theta_1 &= I_0,\\
	\Theta''_3 + \frac{2}{r}\Theta'_3 + (3\omega)^2\Theta_3 &= I_1,\\
	\Theta''_5 + \frac{2}{r}\Theta'_5 + (5\omega)^2\Theta_5 &= I_2,\\
	& \vdots
	\end{cases}
	\label{eq:sinegordon_modes}
\ee
Here, we have introduced the notation
\be \begin{split}
	I_m = \frac{1}{\pi}&\int_0^{2\pi}d\phi \, \cos\((2m+1)\phi\) \times
	\\ & \times \sin\(\sum_n \Theta_{2n+1}(r) \cos\[ (2n+1)\phi\]\).
\end{split} \ee
The set of Eq.~\eqref{eq:sinegordon_modes} is a generalization of Eq.~\eqref{eq:ansatz_wfct} when higher harmonics other than the fundamental mode $\omega$ are considered; when truncating the sum at $n=0$ we obtain the single harmonic approximation Eq.~\eqref{eq:schroedinger_chi} with $\Theta_1 \equiv \Theta$. 

As an example, we consider the case where we also include the first term beyond the single-harmonic approximation besides the fundamental mode $\omega$. This gives
\bea
	\Theta''_1 + \frac{2}{r}\Theta'_1 &=& I_1 - \omega^2\Theta_1, \label{eq:SG_bessel_1}\\
	\Theta''_3 + \frac{2}{r}\Theta'_3 &=& I_3 - 9\omega^2\Theta_3, \label{eq:SG_bessel_2},\\
	I_{2n+1} &\approx& 2(-1)^n J_{2n+1}(\Theta_1) \!+\! \Theta_3\,D_{2n+1}(\Theta_1),
\eea
where we have approximated the computation of the coefficients $I_1$ and $I_3$ by expanding around $\Theta_3 = 0$, with
\be
	D_1(\Theta_1) \approx -\frac{\Theta_1^2}{8},\quad \hbox{and} \quad D_3(\Theta_1) \approx 1-\frac{\Theta_1^2}{4}.
\ee
In fact, the solutions found in Sec.~\ref{non-perturbative solutions} correspond to the zeroth-order approximation of the full non-linear solution, while solving the set of Eqs.~\eqref{eq:SG_bessel_1}--\eqref{eq:SG_bessel_2} gives the next-to-leading order contribution.

\begin{figure}[htbp]
\begin{center}
	\includegraphics[width=\linewidth]{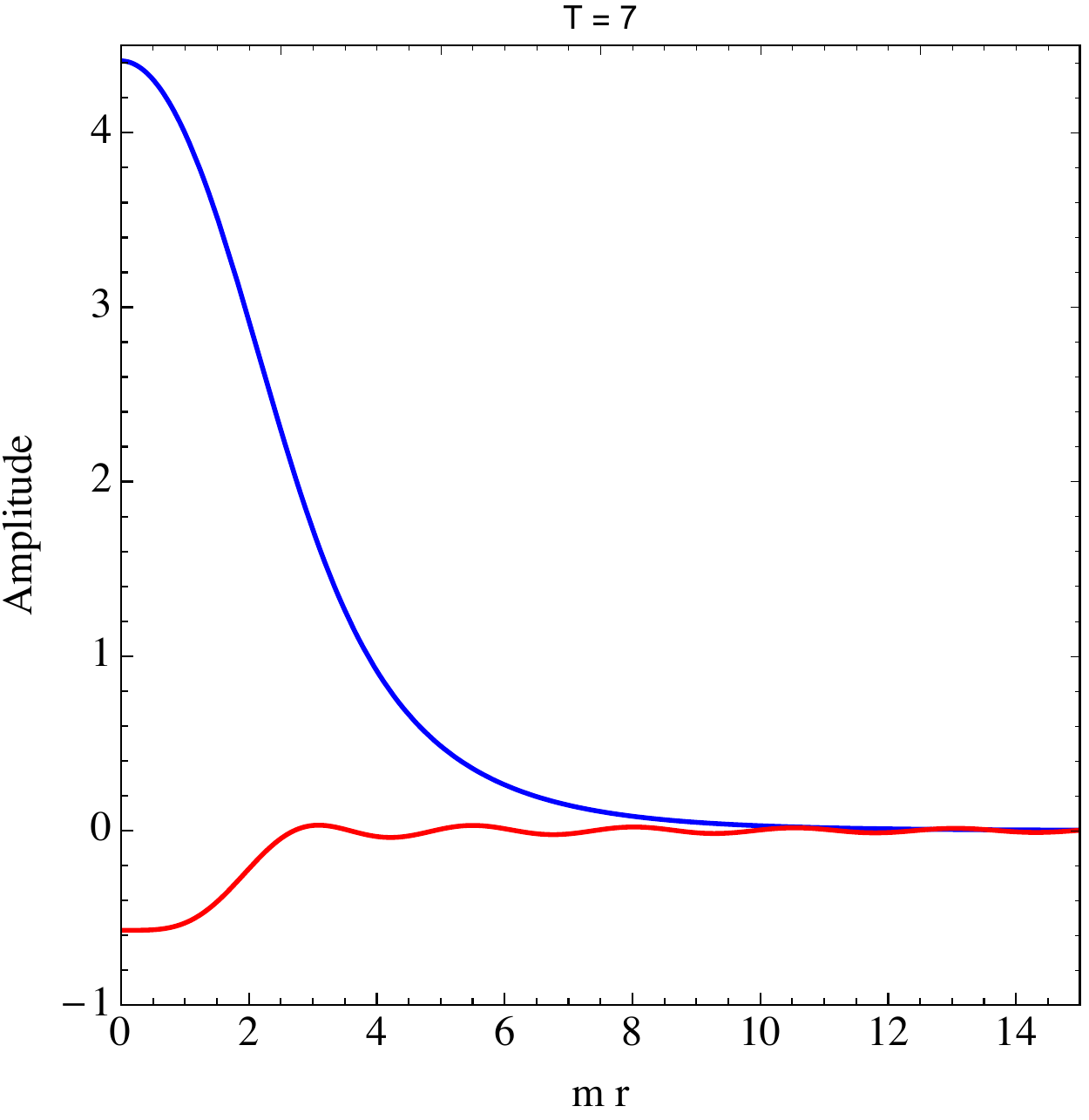}
	\caption{The first harmonic $\Theta_1$ (blue) and the second harmonic $\Theta_3$ (red), satisfying the set of Eqs.~\eqref{eq:SG_bessel_1}-\eqref{eq:SG_bessel_2}, as a function of the radius in units of the axion mass. We fix the axion frequency $\omega = 2\pi/T$, with $T=7$.}
	\label{fig:HHplot}
\end{center}
\end{figure}

At $r\to\infty$, solutions must approach zero, with $\Theta_1, \Theta_3 \ll 1$. In this regime, the higher harmonic $\Theta_3$ must satisfy
\be
	\Theta''_3 + \frac{2}{r}\Theta'_3 + \(9\omega^2-1\)\Theta_3 = -\frac{\Theta_1^3}{24}.
	\label{sucho}
\ee
which, in the range $1/3 < \omega < 1$, is an oscillatory solution in space with wavelength $\sqrt{9\omega^2-1}$. 
Thus, if $\omega > 1/3$, the axion star configuration radiates energy away through the third harmonic, with a contribution that increases with the total energy of the axion $\omega$. Equation \eqref{sucho} can in principle be used to compute the lifetime of the axion star in the dilute branch, where the third harmonics is a very small perturbation of the exact solution. For values of $\omega < 1/3$, there is no radiation solution at $r\to \infty$ and higher harmonics have to be taken into account through Eq.~\eqref{eq:sinegordon_modes}. In Fig.~\ref{fig:HHplot}, we solve the set in Eqs.~\eqref{eq:SG_bessel_1}--\eqref{eq:SG_bessel_2} for a given frequency $\omega = 2\pi/T$, with $T = 7$, using a shooting method to obtain the initial conditions for $\Theta_1$ and $\Theta_3$ at $r = 0$ that satisfy $\Theta_1(+\infty) = \Theta_3(+\infty) = 0$. We stress that the amplitudes of the 1st and 3rd harmonic are of the same order of magnitude everywhere in the star, demonstrating that the single harmonic approximation sufficient for the case of dilute axions stars does not suffice for the description of the dense regime.

\section{Discussion} \label{sec:discussion}

We have shown that when $\Theta_0 \gtrsim \mathcal{O}(1)$, axions are relativistic and axion stars enter the dense branch regime where the configuration behaves as a meta-stable oscillon of the $\chi$-Gordon equation, with a characteristic lifetime. For a free field, bosons stream away from the oscillon core of a star of radius $R$, with a lifetime $\tau_{\rm lin} = 0.836\sqrt{2}R^2$~\cite{Gleiser:2009ys} and with a radiation spectrum peaking at $\omega_{\rm lin}$ with width $\Gamma_{\rm lin} = (2\tau_{\rm lin})^{-1}$. Including a more realistic non-linear self-interaction potential modifies the spectrum by lowering the peak frequency at a lower value $\omega_{\rm nl} < \omega_{\rm lin}$, with a new width $\Gamma_{\rm nl} < \Gamma_{\rm lin}$. Following Ref.~\cite{Gleiser:2009ys}, an oscillon forms if the two spectra do not significantly overlap, that is when
\be
	\omega_{\rm lin} - \omega_{\rm nl} > \frac{\Gamma_{\rm lin} +  \Gamma_{\rm nl}}{2} \approx \Gamma_{\rm lin}.
\ee	
The computation of the oscillon lifetime for a quartic self-interaction has been addressed in Refs.~\cite{Gleiser:2008ty, Gleiser:2009ys}, where the relatively long lifetime (on the scale of the intrinsic timescale $m^{-1}$) of oscillons is explained by the relatively small overlap between the oscillation frequencies.  Following this method, we estimate of the lifetime of an oscillon for a cosine potential as
\be
	\tau_{\rm life} = \frac{1}{\alpha\(E_{\rm osc} - E_{\infty}\)} \approx \frac{700}{m} = 10^{-8}{\rm s},
	\label{eq:lifetime}
\ee
where we have used the parameters $\alpha = 5 \times 10^{-5}$, $E_{\rm osc} = 402.1$, and $E_{\infty} = 372.8$, following Refs.~\cite{Gleiser:2008ty, Gleiser:2009ys} with the axion Lagrangian in Eq.~\eqref{eq:lagrangian} and a Gaussian {\it ansatz\/} for the radial wave function. In short, the energy of an oscillon is described by its radius and amplitude, and damped oscillations in the oscillon develop along the line of constant minimum energy~\cite{Gleiser:1994, Copeland:1995fq}. We performed an independent check of these results by using the solutions of the time-independent Eq.~\eqref{eq:schroedinger_chi} in the dense branch as initial conditions which we time-evolve with the Sine-Gordon Eq.~\eqref{eq:sinegordon}, as prescribed in Ref.~\cite{Piette:1997}. Although this initial wave function is not a proper solution to the Sine-Gordon equation, our numerical solutions yield breather solutions. For a period $T_{\rm nl} = 7.0$ as considered above, we find that the solution decays after $\tau_{\rm life} \approx 1200/m$. This result is of the same order of magnitude as what we obtained using Eq.~\eqref{eq:lifetime}
\be
	\tau_{\rm life} = \mathcal{O}\(\frac{10^3}{m}\) \approx 10^{-7}{\rm \,s} \(\frac{10^{-5}{\rm \,eV}}{m}\). 
\ee
The fact that pseudo-breathers exist has been shown in Ref.~\cite{Hormuzdiar:1999uz}, where the existence of a finite life-time solution to the Sine-Gordon equation has been related to the singular behavior of the solution at zero, when an oscillating function has been imposed as the boundary condition of the solution at infinity. Pseudo-breathers are ultimately decaying states, as discussed in length in Ref.~\cite{Kolb:1993hw}, where it is found numerically that such solutions are unstable and fragment into smaller clumps. The dynamics and the initial conditions considered in Ref.~\cite{Kolb:1993hw} are however different from ours, since the authors consider a cosmological evolution of the axion field with ``white noise'' initial conditions, and included the Hubble rate in the equation of motion.

Recently, Helfer {\it et al.}~\cite{Helfer:2016ljl} have studied the stability of axion stars including gravity and non-linear effects, finding that stable dense profiles may be possible when $f \gtrsim 0.1\,M_{\rm Pl}$, the exact value depending on the axion star mass. In any case, the energy scale $f$ involved is well above the scales we consider here. For values of $f$ below this critical value, the axion star either collapses to a black hole or dissolves by the emission of relativistic particles, consistently with the puffing out obtained in Ref.~\cite{Levkov:2016rkk} and in this work.

\section{Conclusions} \label{sec:conclusion}

In this paper, we have discussed the properties of axion stars for all allowed values of the core amplitude of the axion field $\Theta_0$. In particular, we have discussed how classically stable solutions can arise from the interplay between self-gravity, axion self-interactions, the pressure due to the Heisenberg uncertainty principle, and the kinetic energy. Using assumptions commonly made in the literature, we have obtained a set of equations describing coherent axion field oscillations inside the axion star in the single-harmonic approximation. For small core amplitudes $\Theta_0 \lesssim 1$, we confirmed known results for axion stars in the dilute and critical branches, and provided a heuristic interpretation of those results from first principles. 

For $\Theta \gtrsim 1$, the ``dense'' regime, we recover similar results to those in Ref.~\cite{Braaten:2015eeu} when using the single-harmonic approximation, in particular, the mass radius relation $R \propto M^{1/3}$. However, we argue that the single-harmonic approximation does not hold for the dense regime and thus a different approach is needed, taking into account higher harmonics.  In the end, we arrive a very different  physical interpretation of the dense regime. We find gravity to be negligible for $\Theta = \mathcal{O}(1)$.  Dense axion stars should be solutions to the Sine-Gordon (or $\chi$-Gordon) equation describing the axion field inside the star. We computed the lifetime of dense configurations using both the semi-analytical procedure described in~\cite{Gleiser:1994, Copeland:1995fq, Gleiser:2008ty, Gleiser:2009ys} and by using our single-harmonic solutions as initial conditions, which we time-evolved numerically using the Sine-Gordon equation as prescribed in~\cite{Piette:1997, Alfimov:2000}. Both methods yield comparable lifetimes of order $\tau_{\rm life} \sim 10^3/m$, much shorter than any cosmological time scale.

We conclude that if dense axion stars can be formed, they would immediately (on cosmological scales) radiate relativistic axions and decay.  Since axion stars in the critical branch are unstable against perturbations and either expand to stable dilute configurations or contract to the dense branch and subsequently decay, stable axion stars with mass $M > \tilde{M}^* (f^2/m) \sim 10^{-11}\,M_\odot \left( 10^{-5}\,{\rm eV}/m\right)^2$ appear implausible.

\section*{Additional Note}

During the final preparation of the manuscript after completion of this work we received~\cite{Schiappacasse:2017ham, Chavanis:2017loo}, partially overlapping with this work.

\begin{acknowledgments}
We would like to thank Eric Braaten, Katherine Clough, Malcolm Fairbairn, Oleg Gnedin, Thomas Helfer, \u{Z}elimir Marojevi\'{c}, David J. E. Marsh, and Scott Tremaine for the useful discussions and comments that led to the present work. \\
\phantom{11} SB and LV would like to thank the University of Michigan and the Massachusetts Institute of Technology, where part of this work was conducted, for hospitality.  \\
\phantom{11} SB, KF, and LV acknowledge support by the Vetenskapsr\r{a}det (Swedish Research Council) through contract No. 638-2013-8993 and the Oskar Klein Centre for Cosmoparticle Physics.
 KF acknowledges support from DoE grant DE-SC007859 and the MCTP at the University of Michigan. 
JR is supported by the Ramon y Cajal Fellowship 2012-10597, the grant FPA2015-65745-P (MINECO/FEDER), the EU through the ITN  ``Elusives'' H2020-MSCA-ITN-2015/674896 and the Deutsche Forschungsgemeinschaft under grant SFB-1258 as a Mercator Fellow.
FW's work is supported by the U.S. Department of Energy under grant DE-SC0012567, the European Research Council under grant 742104,
and the Vetenskapsr\r{a}det (Swedish Research Council) under Contract No. 335-2014-7424.
\end{acknowledgments}

\bibliography{axBib}

\end{document}